\documentclass[11pt,a4paper]{article}
\usepackage{amsmath, amscd, amssymb, amsthm, latexsym}
\usepackage{hyperref, textcomp}

\theoremstyle{plain}
\newtheorem{theorem}{Theorem}%[section]

\newtheorem{remark}{Remark}

\theoremstyle{definition}
\newtheorem{definition}[theorem]{Definition}

\usepackage{float}
\usepackage{xcolor}

\newcommand{\zs}{{\mathcal{Z}}}
\newcommand{\ns}{{\mathcal{N}}}

\newcommand{\N}{\mathbb{N}}
\newcommand{\Z}{\mathbb{Z}}
\newcommand{\Q}{\mathbb{Q}}

\newcommand{\mb}[1]{{\mathbf{#1}}}

\def\[#1]{\hbox{$ [\kern -.4em [\, {#1}\, ]\kern -.4em]$}}
\def\[#1]{\hbox{$ [\kern -.4em [\, {#1}\, ]\kern -.4em]$}}
\newcommand{\means}[1]{\hbox{$ [\kern -.4em [\, {#1}\, ]\kern -.4em]$}}

\begin{document} %%%%%%%%%%%%%%%%%%%%%%%%%%%%%%%%%%%%%%%%%%%%%%%%%%%%%%%%%%
%%%%%%%%%%%%%%%%%%%%%%%%%%%%%%%%%%%%%%%%%%%%%%%%%%%%%%%%%%%%%%%%%%%%%%%%%%%

\title{Unary counting quantifiers do not increasing
the expressive power of Presburger arithmetic:
an alternative shorter proof
}

\author{Christian Choffrut\\
IRIF, 
Universit\'{e}  Paris Cit\'{e},
France\\
{\small \tt cc@irif.fr}
}

\date{\today}

\maketitle

This work  was presented  in june 5-7, 2017 at the conference  ``Journ\'{e}es sur les Arithm\'{e}tiques Faibles --
Weak Arithmetics Days'' held in Saint-Pertersburg 
of which no proceeding was ever published. It was not a new result but showed that a different approach is
possible.  The paper presented at ICALP 2024  \cite{haase} addresses, among other problems, the complexity issues
which were ignored in my 2017 talk.

\bigskip
 A \emph{unary counting quantifier}
is a construct of the form $\exists^{=y}_{x}$ and serves as a prefix
of a first order  formula of the Presburger arithmetics, i.e., 
the arithmetics of the integers $\Z$ without the multiplication, denoted $FO(\Z;<,+)$. A
formula $\exists^{=y}_{x_{n}}\phi(x_{1}, x_{2}, \ldots, x_{n})$
is true under the interpretation
$a_{1},a_{2}, \ldots, a_{n-1}$ for $x_{1}, x_{2}, \ldots, x_{n-1}$
and $b$ for $y$
if and only if the number of integer values $a$ 
satisfying $\phi(a_{1}, a_{2}, \ldots, a_{n-1}, a)$ equals  $b$, see \cite[page 646]{schweikardt}. For example the formula 
$\exists^{=y}_{x} (-1\leq x\leq 3)$ interprets to true if and only if $y=5$. 
The logic $FO(\Z;<,+)$ extends to $FOC(\Z;<,+)$ ($c$ for \emph{counting})
by allowing, along with the ordinary quantifiers, those counting quantifiers.
It seems that the term  appeared for the first time in  \cite{barImSt}. However, the idea of introducing 
some kind of counting has former ocurrences. For example 
Apelt \footnote{Apelt refers to H\"{a}rtig for the original 
definition which is equivalent, yet different from that given here.} 
introduced the quantifier $I$ defined as follows:
 the expression $I x (\phi(x) \psi(x))$ holds if the number of values of $x$ satisfying the predicate $\phi$
 equals that satisfying the predicate $\psi$.
 He proved in 1966 
that this logic does not have a greater expressive power 
 than $FO(\Z;<,+)$, \cite[p. 156]{apelt}.   Nicole 
Schweikardt  proves  quantifier elimination  of $FOC(\Z;<,+)$  \cite[Thm 5.4]{schweikardt} 
whose corollary is that adding counting quantifiers does not increase the expressive power of $F0(<,+)$. 
 
 The purpose of this 
work is to give an alternative  proof  using the theory of semilinear sets
and in particular their valuable property that they are finite disjoint unions where counting is easy. It 
can be stated as follows.

\begin{theorem}
\label{th:foc}
Given a Presburger formula
$\phi(x_{1}, \ldots, x_{n})$ with 
free variables $x_{1}, \ldots, x_{n}$, there exists a 
	Presburger formula $\psi(x_{1}, x_{2}, \ldots, x_{n-1},y)$
equivalent to the formula  $\exists^{=y}_{x_{n}} \phi(x_{1}, \ldots, x_{n})$
\end{theorem}
We show how to  use of 
 Ginsburg' and Spanier's 
characterization of Presburger definable
subsets along with the improvement  of  Eilenberg and Sch\"{u}tzenberger,
see the third item of Theorem \ref{th:ginsburg} . We avoid case studies and the application of the inclusion-exclusion principle.

\section{Semilinear sets}

%There are two noticeable families attached with a monoid $(M, \cdot, 1)$ (we refer the reader to the many handbooks 
%on the subject, e.g., \cite{saka}. The family Rat($M$) 
%of {\em rational} 
%subsets it the least family which contains all singletons and which if closed under 
%set union ($X,Y\subseteq M \rightarrow X\cup Y$), set concatenation 
%($X,Y\subseteq M \rightarrow  X\cdot Y= \{x\cdot y\mid x\in X, y\in Y\}$
%and Kleene star (or generated submonoid) 
%($X\subseteq M \rightarrow  X^*= \{1\} \cup x_{1}\cdots x_{n}  \mid  n>0,  x_{1}, \ldots,  x_{n} \in X\}$. 
%
%The family Rec$(M)$ is the family of subsets$X\subseteq M$ for which there exits a monoid 
%morphism $\phi:M\rightarrow N$ where $N$ is a finite monoid, and a subset $S\subseteq N$ such that 
%$X=\phi^{-1}(S)$.
%
%When $M$ is finitely generated we have
%
%\begin{theorem}{\cite{knight} also \cite[Proposition 2.5 ]{saka}}
%Rec($M\subseteq $ Rat($M$).
%\end{theorem}
%
%\begin{theorem}\cite[Proposition 2.6 ]{saka}
%For all $X\in \text{Rec}(M)$ and $Y\in \text{Rat}(M)$ it holds $X\cap Y\in \text{Rat}(M)$.
%\end{theorem}
%
%
%We shall apply these notions to the additive monoids $\Z^n$ or $\N^n$.
%We refer to \cite{eilenberg} for a full exposition of the theory of
%rational subsets of $\N^n$ and $\Z^{n}$. In order to keep our work
%self-contained, we content ourselves with recalling
%the  properties needed for our purpose only. 
%
We view the elements of $\Z^n$ or $\N^n$ as vectors.
The operation of addition  extends  to subsets: if
$X,Y\subseteq\Z^n$, then 
the \emph{sum} $X+Y\subseteq\Z^n$ is the set 
of all sums $\mb{x}+\mb{y}$ where $\mb{x}\in X$ and $\mb{y}\in Y$. 
When $X$ is a singleton $\{\mb{x}\}$ we simply write 
 $\mb{x}+Y$.
Given $\mb{x}$ in  $\Z^n$, the expression $\N \mb{x}$
represents the subset of all vectors $n \mb{x}$ where $n$ ranges 
over $\N$. 
For example, $\N \mb{x}+\N \mb{y}$ 
represents the monoid generated by the vectors $\mb{x}$ 
and $\mb{y}$. 

We need a preliminary definition. 
\begin{definition}
\label{de:semilinear}
A subset of $\Z^{n}$ (resp. $\N^{n}$)  is \emph{linear} 
if it is of the form 
\begin{equation}
\label{eq:semilinear}
\mb{a} + \N \mb{b_{1}} + \cdots + \N \mb{b_{p}}
\end{equation}
for some $n$-vectors $\mb{a}, \mb{b_{1}}, \ldots, \mb{b_{p}}$ in $ \Z^n$
(resp.in $\N^n$). 
It is \emph{simple} if furthermore, the vectors 
$\mb{b_{1}}, \ldots, \mb{b_{p}} $ are linearly independent when considered
as embedded in $\Q^{n}$. 
It is \emph{semilinear} resp. \emph{semisimple} 
if it is a finite union of linear  (resp. simple) sets.
\end{definition}
The main result on semilinear sets is summarized in the Theorem 
below. 
Ginsburg and Spanier proved the 
 equivalence of the first two  statements for $\N^n$ \cite{ginsburg-spanier66},  
but it can readily be seen to hold for $\Z^n$. 
Eilenberg and Sch\"utzen\-berger
 \cite{eilenberg} proved the equivalence of the first two statements
in the general  case of commutative monoids and established 
furthermore their equivalence  
with the last statement  for $\Z$ and $\N$, a result 
which was  left open by Ginsburg and Spanier
and which was independently obtained by Ito \cite{ito}.
\medskip We let  $\zs$ and $\ns$ denote respectively, the first order structure 
$\langle \Z; < , +\rangle$   and $\langle \N;  <, + \rangle$.

\begin{theorem}
\label{th:ginsburg}
Given a subset $X$ of $\Z^{n}$ (resp. $\N^{n}$), the following assertions are equivalent:
%
%(i) $X$ is a rational  subset of $\Z^{n}$ (resp. $\N^{n}$);\\
(i) $X$ is first-order definable in $\zs$ (resp. $\ns$);\\
(ii) $X$ is  $\N$-semilinear;\\
(iii) $X$ is a finite union of disjoint simple subsets.%;\\
%(iv)  $X$ is a rational subset of  $\zs$ (resp. $\ns$).
%\end{enumerate}
\end{theorem}

Consequently, a subset in $\Z^n$ (resp. $\N^n$) is first-order definable in the above structure
if and only if it is a disjoint union of simple subsets of 
$\Z^n$ (resp. $\N^n$).

\section{An  example}

\begin{remark}
\label{re:interval}
Consider the $FO(\Z;<,+)$- predicate  $\gamma_{m,a,D}(y,z,x) = (y\leq mx \leq z \wedge (x=a \bmod D)$ where $m, D, a$ are fixed  parameters $m, D>0$ and $0\leq a<D$. 

The predicate is not satisfiable  if $z<y$. If $y=z$,  it is satisfiable if and only  
if $y=z$ and $x=y  \bmod mD$.

Assume $0<z-y<mD$. If there exists $0\leq i \leq z-y$ such that $y+i= ma\bmod mD$,  it is satisfiable 
if and only if  $x=y+i$.

Assume now $z-y\geq mD$. Set   $y=ma+ i \bmod mD$,  $z=ma + j \bmod mD$ and $0\leq ma +i, ma+j <mD$.  If $i\leq j$ then the formula is true for $\lfloor \frac{z-y}{mD}\rfloor$ different values of $x$ and
otherwise it is true for $\lceil \frac{z-y}{mD}\rceil$ different values for $x$. Formally,  the counting formula $\exists^{u}_{x} \gamma_{m,a,D}(y,z,x)$
is thus equivalent to the $FO(\Z;<, +)$ formula $\delta_{m,a,D}(y,z,u)$
$$
\begin{array}{l}
 (z<y \rightarrow u=0) \wedge (y= z \rightarrow  (y=m a \bmod m D\wedge u=1)) \\
\displaystyle \wedge (y<z) \rightarrow \\
\displaystyle  \big( \bigvee_{-ma\leq i\leq j <-ma + mD} (y=am+i \bmod m D \wedge z=am+ j \bmod mD) \wedge u=\lfloor \frac{z-y}{mD}\rfloor\\
 \displaystyle \vee \bigvee_{-ma\leq j< i<-ma +mD} (y=am+i \bmod m D \wedge z=am+ j \bmod mD) \wedge u=\lceil \frac{z-y}{mD}\rceil\big)
\end{array}
$$

\end{remark}

We study an  example in order to highlight  the specific 
properties that we take advantage of in order to more easily produce an equivalent ordinary  Presburger predicate 
to a given  
predicate with counting quantifier. Consider the first-order formula (with the notations of (\ref{eq:semilinear}) the vector $\bf{a}$ is null, $p=3$ and the vectors  $\bf{b}_{1}, \bf{b}_{2}$ and $\bf{b}_{3}$
are reepectvely $(1,2,2,1)^{T}$, $(2,4,1,1)^{T}$ and  $(-1, -2, 0, -1)^{T}$)
$$
\begin{array}{l}
\phi(x_{1}, x_{2}, x_{3},x_{4}) \equiv  \exists z_{1}, z_{2}, z_{3}:  z_{1}, z_{2}, z_{3}\geq 0\\
(x_{1} = z_{1} + 2z_{2} - z_{3}) \wedge
(x_{2} = 2z_{1} +  4z_{2}    -  2z_{3})  \wedge
(x_{3}  =2z_{1} + z_{2} )\wedge
(x_{4}  =z_{1} +  z_{2}   - z_{3} )
\end{array}
$$
which we write as a  system of linear equations
$$
\begin{array}{llllll}
z_{1} &+ &2z_{2} & - & z_{3} &= x_{1} \\
2z_{1} &+ & 4z_{2} &   - & 2z_{3} & = x_{2} \\
2z_{1} &+ &z_{2} & &  & = x_{3}\\
z_{1} &+ & z_{2} & - & z_{3} &= x_{4}\end{array}
$$
%
%The two specific features enjoyed by this example is the lack of 
%disjunction and the fact that every submatrix of rank 3 necessarily 
%contains the  row of the matrix corresponding to the variable
%to be counted, namely the fourth row {\color{red} (explain)}. We will see that these two conditions can always be assumed.  
The subsystem consisting of the first, third and fourth rows has determinant equal to $2$. We solve 
the subsystem in the unknowns $z_{1}$, $z_{2}$ and $z_{3}$, which yields
$$
\begin{array}{ll}
2 z_{1} & = - x_{1} +   x_{3} +x_{4}\\
2 z_{2} & =  2 x_{1} - 2 x_{4} \\
2 z_{3} & = x_{1} +  x_{3}  - 3 x_{4}\\
\end{array}
$$
Now, we must express the fact that 
the variables $z_{1}, z_{2}, z_{3}$ are positive integers. This is the case 
if and only if the following conditions hold (the coefficient $6$
is the positive least common multiple of the coefficients of the variable $x_{4}$)

\begin{align}
6 x_{4} & \geq  6  x_{1} -   6 x_{3} \nonumber  \\
 6 x_{4} & \leq  6 x_{1} \nonumber \\
6 x_{4} &  \leq   2 x_{1} +  2 x_{3}    \nonumber\\
 x_{1} +   x_{3} +x_{4} & =0 \bmod 2 \label{eq:modular}
\end{align}
The first three conditions are  equivalent  to
\begin{equation}
\label{eq:which-is-max}
6  x_{1} -   6 x_{3} \leq 6 x_{4} \leq   \min \{  6 x_{1} , 2 x_{1} +  2 x_{3} \} 
\end{equation}
There are four different cases according to whether or not
$ 2 x_{1} +  2 x_{3} \leq 6  x_{1} $ and whether or not $x_{1} +   x_{3}=0\bmod 2$.
For example if these two conditions hold, implying in particular 
because of    (\ref{eq:modular}) that  $x_{4}=0\bmod 2$, we are led  consider the number of 
even integers  satisfying the condition 
\begin{equation}
\label{eq:ex-interval}
6  x_{1} -   6 x_{3} \leq 6 x_{4} \leq   2 x_{1} +  2 x_{3}  
\end{equation}
which can be done following the lines of  Remark \ref{re:interval} with $\delta_{2,0,6}(6  x_{1} -   6 x_{3},  2 x_{1} +  2 x_{3}, u)$.

\section{The proof}

Because of  item $(iii)$ of Theorem \ref{th:ginsburg}, every
formula of Presburger arithmetic with free variables 
$x_{1}, \ldots, x_{n}$ is equivalent 
to a formula of the form 
$$
\phi(x_{1}, \cdots, x_{n}) 
\equiv \phi_{1}(x_{1}, \cdots, x_{n})\vee \cdots \vee \phi_{r}(x_{1}, \cdots, x_{n})
$$
where the  $\phi_{i}$'s define disjoint  simple subsets of $\Z^n$
which implies
$$
\begin{array}{l}
\exists^{=y}_{x_{n}}  \phi(x_{1}, \cdots, x_{n})\equiv \exists y_{1}, \ldots, \exists y_{r} \\
\big( \exists^{=y_{1}}_{x_{n}} \phi_{1}(x_{1}, \cdots, x_{n})\vee \cdots 
\vee \exists^{=y_{r}}_{x_{n}}  \phi_{r}(x_{1}, \cdots, x_{n})\big)
\wedge (y_{1}+ \cdots + y_{r}=y)
\end{array}
$$
It thus suffices to prove the case  $r=1$, which means 
that we can assume that $\phi(x_{1}, \cdots, x_{n})$
defines a simple subset. We  express
the problem in terms  of linear algebra. 
We use  the expression (\ref{eq:semilinear})
and we let $M\in \Z^{n\times p}$  denote the matrix
of rank $p$ whose columns are the linearly independent vectors 
$\mb{b_{1}}, \cdots, \mb{b_{p}}$. We are 
interested in solving the following equation
where $\mb{x}$ and $\mb{a}$ are $n$-column integer vector and 
$\mb{z}$ is a $p$-column nonnegative integer vector
\begin{equation}
\label{eq:matrix}
 \mb{a} +  M \mb{z}=\mb{x}
\end{equation}
In particular we get
$$
\phi(\mb{x}) \Leftrightarrow \exists  \mb{z}\in \N^{p}:   \mb{a}+  M \mb{z}=\mb{x}
$$
%
%\break
%
With the convention that
$b_{i,j}$ and $a_{i}$ are the $i$-th components of the vector $\mb{b}_{j}$
and $\mb{a}$ respectively, 
this is equivalent to the following system of equations
\begin{equation}
\label{eq:linear-system}
\begin{array}{lllcll}
b_{1,1} z_{1} &+& \cdots &+& b_{1,p} z_{p} &= x_{1} -a_{1} \\
 \ldots & & & \\
b_{n,1} z_{1} &+& \cdots &+& b_{n,p} z_{p} &=x_{n}  - a_{n} 
\end{array}
\end{equation}
The matrix has rank $p\leq n$.
If there is a submatrix of rank $p$ obtained by selecting
$p$ among the $n-1$ first rows, then the $x_{i} -a_{i}$'s
for which $i$ is the index of a row among the 
selected rows, define uniquely all $x_{j} -a_{j}$'s
for all indices  outside the selected rows.
In particular there is a unique possible value for $x_{n} -a_{n}$'s.
A Presburger formula expressing this 
relation is 
$$
\exists^{=y}_{x_{n}} \phi(x_{1}, \ldots, x_{n}) \equiv \exists x_{n} \phi(x_{1}, \ldots, x_{n}) \wedge y=1.
$$

Consider now the second case where all submatrices of 
rank $p$ contain the last row. This means that there exist $p-1$  among the 
 $n-1$ first rows that determine the values of the variables $x_{i}$, for $i<n$. Thus we may assume without loss of generality 
 that $n=p$.

 By Cramer's rules,  $z_{1}, \ldots
z_{p}$ can be uniquely expressed as a function  of $x_{i}$'s, i.e., 
\begin{equation}
\label{eq:linear-expression}
\begin{array}{l}
\displaystyle D z_{i}= \lambda_{i,p} x_{p}  + \sum^{p-1}_{j=1}\lambda_{i,j} x_{j}  +  \gamma_{i} \quad i \in \{1, \ldots, p\}
\end{array}
\end{equation}
where $D$ is  the absolute value of the determinant of the matrix $M$
and where the coefficients $\lambda_{i,j},  \gamma_{i}$ are integers.
We want to express in FO$(\Z;<, +)$ the fact that the $z_{i}$'s are nonnegative integers. 
To that purpose let $F$ be the set of mappings 
$f:\{1, \ldots, p-1\}\mapsto  \{0, \ldots, D-1\}$. Then $\phi$ is equivalent to the disjunction, over all functions 
$f\in F$, of the following predicates $\phi^{(f)}(x_{1}, \ldots, x_{n}) $
$$
\phi^{(f)}(x_{1}, \ldots, x_{n})  \equiv \phi(x_{1}, \ldots, x_{n})  \displaystyle %\sum_{j} \lambda_{i,j} f(j)  + \gamma_{i} \equiv_{D} 0
\wedge \bigwedge_{1\leq j<n} x_{j} = f(j) \bmod D) \wedge ( \bigvee_{0\leq a< D}  x_{p} = a  \bmod D)
$$
If $\phi^{(f)}(x_{1}, \ldots, x_{n})  $  is not satisfiable then neither is
$\exists^{=y}_{x_{n}}\phi^{(f)}(x_{1}, \ldots, x_{n})$. 
Observe that the relations defined  when $f$ ranges over $F$ and $a$ over
$\{0, \ldots, D-1\}$ are disjoint. Thus we concentrate on a specific $f$ and a specific $a$.
$$
\psi(x_{1}, \ldots, x_{n}) =  \displaystyle  (\bigwedge_{1\leq j<n} x_{j} = f(j) \bmod D) \wedge (  x_{p} = a  \bmod D)
$$ 
We construct an $FO(\Z;<,+)$ formula equivalent to 
$\exists^{=u}_{x_{n}} \psi(x_{1}, \ldots, x_{n})$.
We  set $-\lambda_{i,p}=\frac{m}{\eta_{i}}$ where $m$ is the least common positive multiple
of the nonzero $\lambda_{i,p}$'s and we let   $S_{i}(x_{1}, \ldots, x_{p-1})$ be the polynomial
$ \sum^{p-1}_{j=1}\lambda_{i,j} x_{j}  +  \gamma_{i}$. Henceforth, in order to alliviate the notations we let
the bold face $\bf{y}$ and $\bf{x}$ denote the vectors $(x_{1}, \ldots, x_{p-1})$ and 
$(x_{1}, \ldots, x_{p})$ respectively. 
We set. 
$$
\left\{
\begin{array}{ll} 
\displaystyle U_{i}({\bf{y}}) =\eta_{i} S_{i} & \text{if } \eta_{i}  > 0\\
\displaystyle E_{i}({\bf{y}}) =  S_{i} & \text{if } \lambda_{i,p}= 0\\
\displaystyle L_{i}({\bf{y}}) = \eta_{i}  S_{i}
& \text{if } \eta_{i} <0
\end{array}
\right.
$$

Let $A\subseteq \{1, \ldots, p\}$
be  the set of  indices $i$ for which  $\eta_{i} > 0$ and let
 $B\subseteq \{1, \ldots, p\}$ be  the set of indices $i$ for which  $\eta_{i} < 0$. 
Then,  the  $z_{i}$'s are nonnegative integers if and only if the following holds
\begin{align} 
U_{i}(\bf{y}) & \geq m x_{p}    \text{ for all }  i\in A \label{eq:upper} \\
E_{i}(\bf{y}) & \geq 0    \text{ for all  }  i\not\in A\cup B \label{eq:equal} \\
L_{i}(\bf{y})&    \leq m x_{p}   \text{ for all }  i\in B \label{eq:lower}% \\
\end{align}
If $A=\emptyset$, for a fixed interpretation $a_{1}, \ldots, a_{p-1}$
of the variables  $\bf{y}$ satisfying all predicates $E_{!}\geq 0$, $i\notin A\cup B$, 
there are infinitely 
many  positive values $b$ satisfying all $L_{i}\leq m x_{p}$, $i\in B$  thus also  $\psi(a_{1}, \ldots, a_{p-1},b)$.
 By convention 
we set $\exists^{=y}_{x_{p}}\psi = \texttt{false}$ and we treat similarly the case where
$B=\emptyset$. We thus assume $A, B\not=\emptyset$ with $r$ elements in $A$ and $s$ in $B$.
We set
$$
\begin{array}{l}
\label{eq:}
	\displaystyle \+E(\mb{y})=\bigwedge_{k\notin A\cup B} E_{\gamma}(\mb{y})\geq 0 
\end{array}
$$
We define for all permutations $\sigma$ and $\tau$ of $\{1, \ldots, r\}$ and $\{1, \ldots, s\}$
\begin{equation}
\label{eq:L-R-chains}
\left\{
\begin{array}{l}
	\displaystyle\+L_{\tau}(\mb{y}) \equiv  \bigwedge_{s\geq i >1}  L_{\tau(i)}(\mb{y})< L_{\tau(i-1)}(\mb{y})\\
	\displaystyle\+U_{\sigma}(\mb{y}) \equiv \bigwedge_{1\leq i<r} U_{\sigma(i)}(\mb{y})< U_{\sigma(i+1)}(\mb{y})\\
		\displaystyle \psi_{\sigma,\tau}(\mb{x}) \equiv \+E(\mb{y})\wedge \big( \+L_{\tau}(\mb{y}) \wedge \+U_{\sigma}(\mb{y}) \rightarrow   L_{\tau(1)}(\mb{y})\leq m x_{p}\leq U_{\sigma(1)}(\mb{y}) \big)\\
\end{array}
\right.
\end{equation}
Then the relation defined by $\psi(\bf{x})$ is the disjoint union of the relations defined by the different 
$\psi_{\sigma,\tau}(\mb{x})$,  which yields
$$
\begin{array}{l}
\label{eq:} 
\exists^{=u}_{x_{p}}\psi(\mb{x})\equiv 	\displaystyle  \big(\bigvee_{\sigma, \tau}
	\exists^{=u_{\sigma,\tau}}_{x_{p}} \psi_{\sigma,\tau}(\mb{x}) \big) \wedge \displaystyle\sum_{\sigma, \tau} 
	u_{\sigma,\tau} = u\
\end{array}
$$
Therefore with the notations of Remark \ref{re:interval}, the expression
$\exists^{=u_{\sigma,\tau}}_{x_{p}} \psi_{\sigma,\tau}(\mb{x}) $ is equivalent to the following $FO(\Z;<,+)$ formula
$$
\begin{array}{l}
\label{eq:last} 
\+E(\mb{y})\wedge  \+L_{\tau}(\mb{y}) \wedge \+U_{\sigma}(\mb{y}) \wedge  
 \delta_{m,a,D}( L_{\tau(1)}(\mb{y}),U_{\sigma(1)}(\mb{y}), u_{\sigma,\tau} )
\end{array}
$$

\section{The structure $\ns$}

%An example
%
%$$
%\begin{array}{llllll}
%z_{1} &+ &2z_{2} & + & z_{3} &= x_{1} \\
%2z_{1} &+ & 4z_{2} &   + & 2z_{3} & = x_{2} \\
%2z_{1} &+ &z_{2} & &  & = x_{3}\\
%z_{1} &+ & z_{2} & + & z_{3} &= x_{4}\end{array}
%$$
%%
%The subsystem consisting of the first, third and fourth rows has determinant equal to $2$. We solve 
%the subsystem in the unknowns $z_{1}$, $z_{2}$ and $z_{3}$, which yields
%%
%$$
%\begin{array}{ll}
%2 z_{1} & = - \frac{1}{2} x_{1} + \frac{1}{2} x_{3}. + \frac{1}{2} x_{4} \\
%2 z_{2} & =  2 x_{1} - 2 x_{4} \\
%2 z_{3} & = x_{1} +  x_{3}  - 3 x_{4}\\
%\end{array}
%$$

The task consists essentially in transforming  the linear equalities $A=B$ and inequalities such as $A>B$  
by shifting the monomials with nonnegative coefficients to the  other side  of the sign $=$ or $>$, e.g., 
$x-3y= y-z$ is transformed  into $x+z = 4y$.
\begin{remark}
\label{re:interval-N}
Consider the $FO(\N;<,+)$ predicate  
$$\gamma^{\N}_{m,a, D}(y_{1}, y_{2}, ,z_{1}, z_{2},x) = (y_{1} \leq mx + y_{2}) \wedge (z_{1} +  mx\leq  z_{2}) \wedge (x =a \bmod D)
$$ 
where $m, D>0$ and $0\leq a <D$. There exists a $FO(\N;<,+)$ formula 
$\delta^{\N}_{m,D,a}(y_{1}, y_{2}, ,z_{1}, z_{2},u)$ which is equivalent to the $FOC(\N;<,+)$ formula
 $\exists^{=u}_{x}\phi(y_{1}, y_{2}, z_{1}, z_{2},x)$. 

Indeed, if $z_{1}>z_{2}$ the predicate is not satisfiable. So we have $ mx\leq  z$ with $z= z_{2}-z_{1}\geq 0$. 
Assume first  $y_{2}> y_{1}$. The predicate reduces to $ mx\leq  z$. If $z<mD$ then the predicate is satisfiable if and only if
$y=mx$. Otherwise let $z=ma+i \bmod mD$ with $ma+i \leq mD$. If $i\geq 0$ then the number of values for $x$
equals $\lceil \frac{z}{mD}\rceil $, else it equals $\lfloor\frac{z}{mD}\rfloor $
Now assume  $y_{2}\leq  y_{1}$.  By posing $y=y_{2}-  y_{1}$ the predicate reduces to $y\leq mx\leq  z$ and it suffices to proceed as in remark \ref{re:interval}.

\end{remark}

As explained above we transform every  inequality $E_{i}({\bf{y}}) \geq  0 $ in \ref{eq:equal}
into $E^{(1)}_{i}({\bf{y}}) \geq  E^{(2)}_{i}(\bf{y}) $and we put
$$
\+E^{\N}(\mb{y})= \displaystyle \bigwedge_{i}  E^{(1)}_{i}({\bf{y}}) \geq  E^{(2)}_{i}(\bf{y})
$$
Similarly, we transform 
  $ L_{\tau(i)}(\mb{y})< L_{\tau(i-1)}(\mb{y})$ for $i=s, \ldots, 2$ 
into  $ L^{\N}_{\tau(i)}(\mb{y})< L^{\N}_{\tau(i-1)}(\mb{y})$ and 
 $U_{\sigma(i)}(\mb{y})< U_{\sigma(i+1)}(\mb{y})$  for $i=1, \ldots, r-1$  into $U^{\N}_{\sigma(i)}(\mb{y})< U^{\N}_{\sigma(i+1)}(\mb{y})$.
Also, we transform  inequality $L_{\tau(1)}(\mb{y})\leq mx_{p}$
into $L^{(1)}_{\tau(1)}(\mb{y})\leq mx_{p} + L^{(2)}_{\tau(1)}(\mb{y})$
and $mx_{p}\leq U_{\sigma(1)}(\mb{y})$
into $U'^{(1)}_{\sigma(1)}(\mb{y})\leq mx_{p} + U'^{(2)}_{\sigma(1)}(\mb{y})$.
Applying these transformations to \ref{eq:L-R-chains} we obtain
$$
\begin{array}{ll}
\label{eq:}
	\displaystyle\+L^{\N}_{\tau}(\mb{y}) \equiv & \bigwedge_{s\geq i >1}  L^{\N}_{\tau(i)}(\mb{y})< L^{\N}_{\tau(i-1)}(\mb{y})\\
	\displaystyle\+U^{\N}_{\sigma}(\mb{y}) \equiv &\bigwedge_{1\leq i<r} U^{\N}_{\sigma(i)}(\mb{y})< U^{\N}_{\sigma(i+1)}(\mb{y})  \\
\\
		\displaystyle \psi^{\N}_{\sigma,\tau}(\mb{x}) \equiv &\+E^{\N}(\mb{y})\wedge \big( \+L^{\N}_{\tau}(\mb{y}) \wedge \+U^{\N}_{\sigma}(\mb{y}) \rightarrow  \\
		&  (L^{(1)}_{\tau(1)}(\mb{y})\leq mx_{p} + L^{(2)}_{\tau(1)}(\mb{y})) \wedge U'^{(1)}_{\sigma(1)}(\mb{y})\leq mx_{p} + U'^{(2)}_{\sigma(1)}(\mb{y}) \big)\\
\end{array}
$$
Then the relation defined by $\psi^{\N}(\bf{x})$ is the disjoint union of the relations defined by the different 
$\psi^{\N}_{\sigma,\tau}(\mb{x})$,  which yields
$$
\begin{array}{l}
\label{eq:} 
\exists^{=u}_{x_{p}}\psi^{\N}(\mb{x})\equiv 	\displaystyle  \big(\bigvee_{\sigma, \tau}
	\exists^{=u_{\sigma,\tau}}_{x_{p}} \psi^{\N}_{\sigma,\tau}(\mb{x}) \big) \wedge \displaystyle\sum_{\sigma, \tau} 
	u_{\sigma,\tau} = u\
\end{array}
$$
Therefore with the notations of Remark \ref{re:interval-N}, the expression
$\exists^{=u_{\sigma,\tau}}_{x_{p}} \psi^{\N}_{\sigma,\tau}(\mb{x}) $ is equivalent to the following $FO(\N;<,+)$ formula
$$
\begin{array}{l}
\label{eq:last} 
\+E(^{\N}\mb{y})\wedge  \+L^{\N}_{\tau}(\mb{y}) \wedge \+U^{\N}_{\sigma}(\mb{y}) \wedge  
 \delta^{\N}_{m,D,a}((L^{(1)}_{\tau(1)}, L^{(2)}_{\tau(1)}(\mb{y})),  U'^{(1)}_{\sigma(1)},  U'^{(2)}_{\sigma(1)}, u_{\sigma,\tau} )
\end{array}
$$
which completes the proof.

%%%%%%%%%%%%%%%%%%%%%%%%%%%%%%%%%%%%%%%%%%%%%%%%%%%%%%%%%%%%%%%%%%%%%%%%%%%
\end{document}